\documentclass[12pt]{article}
\usepackage{epsfig}
\usepackage{amsfonts}
\begin{document}
\begin{titlepage}
\begin{center}

{\Large Dynamical foundations of nonextensive statistical mechanics}

\vspace{2.cm} {\bf Christian Beck}\footnote{
permanent address: School of Mathematical Sciences, Queen Mary,
University of London, Mile End Road, London E1 4NS.}

\vspace{0.5cm}

Isaac Newton Institute for Mathematical Sciences,
University of Cambridge, 20 Clarkson Road, Cambridge CB3 0EH, UK

\vspace{5cm}

\end{center}

\abstract{We construct classes of stochastic differential equations
with fluctuating friction forces
that generate a dynamics correctly described by Tsallis statistics
and nonextensive statistical mechanics. These systems generalize
the way in which
ordinary Langevin equations underly
ordinary statistical mechanics to the more general nonextensive case.
As a main example, we
construct a dynamical model of velocity fluctuations in a turbulent
flow, which generates probability densities that
very well fit
experimentally measured probability densities in Eulerian and Lagrangian
turbulence.
Our approach provides
a dynamical reason why many physical systems with fluctuations in
temperature or energy dissipation rate are correctly
described by Tsallis statistics.}

\vspace{1.3cm}

\end{titlepage}

Recently there has been considerable interest in the formalism
of nonextensive statistical mechanics (NESM) as introduced by Tsallis
\cite{tsa1} and further developed by many others (e.g.\ \cite{abe}--\cite{tsa2}).
In the mean time there is growing evidence that the formalism, rather than being
just a theoretical construction, is of
relevance to many complex physical systems.
Applications in various
areas have been reported, mainly for systems
with either long-range interactions \cite{plastino}--\cite{ruffo}, multifractal
behaviour \cite{lyra,ari}, or fluctuations of temperature or
energy dissipation rate \cite{wilk}--\cite{e+-2}.
A recent interesting application of the formalism is that
to fully developed turbulence \cite{ari,hydro,BLS}. Precision measurements
of probability density functions (pdfs) of longitudinal
velocity differences in high-Reynolds number
turbulent Couette-Taylor flows are found
to agree quite perfectly with analytic formulas of pdfs as
predicted by NESM \cite{BLS}.

Despite this apparent success of the nonextensive approach,
still the question remains {\em why} in many
cases (such as the above turbulent flow) NESM works so well.
To answer this question, let us first go back to ordinary
statistical mechanics and just consider a very simple well known
example, the Brownian particle \cite{vKa}.
Its velocity $u$
satisfies the linear Langevin equation
\begin{equation}
\dot{u}=-\gamma u +\sigma L(t), \label{1}
\end{equation}
where $L(t)$ is Gaussian white noise, $\gamma >0$ is a friction
constant, and $\sigma$ describes the strength of the noise.
The stationary probability density of $u$ is Gaussian with average 0
and variance $\beta^{-1}$, where
where
$\beta=\frac{\gamma}{\sigma^2}$
can be identified with the inverse temperature of ordinary
statistical mechanics (we assume that the Brownian particle
has mass 1).

The above simple situation completely changes if one
allows
the parameters $\gamma$ and $\sigma$ in
the stochastic differential equation (SDE) to fluctuate as well.
To be specific, let us assume that either $\gamma$ or $\sigma$ or both
fluctuate in such a way that $\beta=\gamma/\sigma^2$ is $\chi^2$-distributed
with degree $n$. This means
the probability density of $\beta$
is given by
\begin{equation}
f (\beta) = \frac{1}{\Gamma \left( \frac{n}{2} \right)} \left\{
\frac{n}{2\beta_0}\right\}^{\frac{n}{2}} \beta^{\frac{n}{2}-1}
\exp\left\{-\frac{n\beta}{2\beta_0} \right\} \label{fluc}
\end{equation}
The $\chi^2$ distribution is a typical distribution that naturally
arises in many circumstances. For example, consider $n$ independent Gaussian
random variables $X_i,\;i=1,\ldots ,n$ with average $0$. If $\beta$ is
given by
the sum
\begin{equation}
\beta:=\sum_{i=1}^{n} X_i^2 \label{Gauss}
\end{equation}
then it has the pdf
(\ref{fluc}). The average is given by
\begin{equation}
\langle \beta \rangle =n\langle X^2\rangle=\int_0^\infty\beta f(\beta) d\beta= \beta_0
\end{equation}
and the variance by
\begin{equation}
\langle \beta^2 \rangle -\beta_0^2=
\frac{2}{n} \beta_0^2
\end{equation}
(see also \cite{wilk}).

Now assume that the time scale on which $\beta$ fluctuates
is much larger than the typical time scale of
order $\gamma^{-1}$ that the Langevin system (\ref{1})
needs to reach equilibrium.
%In other words, we assume that the system
%reaches the equilibrium state
%fast enough.
In this case one obtains
for the conditional probability $p(u|\beta)$ (i.e. the probability
of $u$ given some value of $\beta$)
\begin{equation}
p(u|\beta)=\sqrt{\frac{\beta}{2\pi}}\exp \left\{ -\frac{1}{2}\beta u^2\right\},
\end{equation}
for the joint probability $p(u,\beta)$ (i.e. the probability
to observe both a certain value of $u$ and a certain value of $\beta$)
\begin{equation}
p(u,\beta)=p(u|\beta)f(\beta)
\end{equation}
and for the marginal probability $p(u)$ (i.e.\ the probability
to observe a certain value of $u$ no matter what $\beta$ is)
\begin{equation}
p(u)=\int p(u|\beta)f(\beta)d\beta . \label{9}
\end{equation}
The integral (\ref{9}) is easily evaluated and one obtains
\begin{equation}
p(u)=\frac{\Gamma\left(\frac{n}{2}+\frac{1}{2}\right)}{ \Gamma
\left(\frac{n}{2} \right)} \left( \frac{\beta_0}{\pi n} \right)^{\frac{1}{2}}
\frac{1}{\left( 1+\frac{\beta_0}{n}u^2 \right)^{\frac{n}{2}+\frac{1}{2}}}
\end{equation}
Hence the SDE
(\ref{1}) with $\chi^2$-distributed $\beta=\gamma/\sigma^2$
generates the generalized canonical
distributions of NESM \cite{tsa1}
\begin{equation}
p(u) \sim \frac{1}{\left( 1+\frac{1}{2}\tilde{\beta}(q-1)u^2\right)^{\frac{1}{q-1}}}
\end{equation}
provided
the following identifications are made.
\begin{eqnarray}
\frac{1}{q-1}=\frac{n}{2}+\frac{1}{2}
&\Longleftrightarrow& q=1+\frac{2}{n+1} \\
\frac{1}{2}(q-1)\tilde{\beta}= \frac{\beta_0}{n}
&\Longleftrightarrow& \tilde{\beta}=\frac{2}{3-q} \beta_0.
\end{eqnarray}
We already see from this simple
example that the physical inverse temperature $\beta_0=\langle \beta \rangle$
not necessarily coincides with the inverse temperature
parameter $\tilde{\beta}$ used in the nonextensive formalism
(see also \cite{abe2} for related results).

More generally, we may also consider nonlinear Langevin equations
of the form
\begin{equation}
\dot{u}=-\gamma F(u)+\sigma L(t) \label{there}
\end{equation}
where $F(u)=-\frac{\partial}{\partial u}V(u)$ is a nonlinear forcing.
To be specific, let us assume that
$V(u)=C|u|^{2\alpha}$ is a power-law
potential.
The SDE (\ref{there})
then generates the conditional pdf
\begin{equation}
p(u|\beta)=\frac{\alpha}{\Gamma \left( \frac{1}{2\alpha}\right)}
\left( C\beta \right)^\frac{1}{2\alpha}\exp\left\{-
\beta C |u|^{2\alpha}\right\}
\end{equation}
and for the marginal distributions $p(u)=\int p(u|\beta)f(\beta)d\beta$
we obtain after a short calculation
\begin{equation}
p(u)=\frac{1}{Z_q}\frac{1}{(1+(q-1)\tilde{\beta}C
|u|^{2\alpha})^{\frac{1}{q-1}}} \label{pu},
\end{equation}
where
\begin{equation}
Z_q^{-1}= \alpha \left( C(q-1)\tilde{\beta}\right)^{\frac{1}{2\alpha}}
\cdot \frac{\Gamma \left( \frac{1}{q-1}\right)}{ \Gamma 
\left( \frac{1}{2\alpha} \right) \Gamma \left(
\frac{1}{q-1}-\frac{1}{2\alpha} \right)}
\end{equation}
and
\begin{eqnarray}
q&=&1+\frac{2\alpha}{\alpha n+1} \label{qn} \\
\tilde{\beta}&=&\frac{2\alpha}{1+2\alpha -q}\beta_0.
\end{eqnarray}

To generalize to $N$ particles in $d$ space dimensions, we may consider
coupled systems of SDEs with fluctuating
friction forces, as given by
\begin{equation}
\dot{\vec{u}}_i=-\gamma_i \vec{F}_i(\vec{u}_1,\ldots ,
\vec{u}_N)+\sigma_i {\vec L}_i(t) \;\;\;\;\;\;\;i=1,\ldots , N
\label{general}
\end{equation}
Suppose that a potential $V(\vec{u_1},\ldots ,\vec{u_N})$ exists
for this problem such that $\vec{F}_i=\frac{\partial}{\partial \vec{u}_i}V$.
Moreover, assume
that
all $\beta_i=\frac{\gamma_i}{\sigma^2_i}$ fluctuate in the same way,
i.e. are given by the same fluctuating $\chi^2$
distributed random variable $\beta_i=\beta$.
One then has for the conditional probability
\begin{equation}
p(\vec{u}_1,\ldots ,\vec{u}_N|\beta)
=\frac{1}{Z(\beta)}\exp{\left\{ -\beta V(\vec{u}_1,\ldots ,\vec{u}_N)
\right\}},
\end{equation}
where $Z(\beta)=\int d\vec{u}_1\ldots d\vec{u}_N\;
e^{-\beta V}$ is the partition function of ordinary statistical
mechanics. Suppose
that $Z(\beta)$ is of the form
\begin{equation}
Z(\beta) \sim \beta^x e^{-\beta y},
\end{equation}
then integration over the fluctuating $\beta$ leads to marginal distributions
of the form
\begin{equation}
p(\vec{u_1},\ldots ,\vec{u_N})\sim \frac{1}{(1+\tilde{\beta}(q-1)
V(\vec{u_1},\ldots ,\vec{u}_N))^\frac{1}{q-1}} ,\label{dense}
\end{equation}
i.e. the generalized canonical distributions of NESM with
\begin{equation}
q=1+\frac{2}{n-2x}
\end{equation}
and
\begin{equation}
\tilde{\beta}=\frac{\beta_0}{1+(q-1)(x-\beta_0 y)}.
\end{equation}

Eq.~(\ref{dense}) is correct if all particles see the same
fluctuating $\beta$ at the same time---an assumption that can only
be true for a very dense and concentrated system of particles. In
many physical applications, the various particles will be dilute
and only weakly interacting. Hence in this case $\beta$ is
expected to fluctuate spatially in such a way that the local
inverse temperature $\beta_i$ surrounding one particle $i$ is
almost independent from the local $\beta_j$ surrounding another
particle $j$. Moreover, the potential is approximately just a sum
of single-particle potentials $V(\vec{u}_1,\ldots ,\vec{u}_N)
=\sum_{i=1}^NV_s(\vec{u}_i)$.
In this case integration over all
$\beta_i$ leads to marginal densities of the form
\begin{equation}
p(\vec{u}_1,\ldots ,\vec{u}_N)\sim \prod_{i=1}^N \frac{1}{(1+\tilde{\beta}(q-1)
V_s(\vec{u}_i))^\frac{1}{q-1}}, \label{dilute}
\end{equation}
i.e.\ the $N$-particle nonextensive system reduces to products of
1-particle nonextensive systems (this type of factorization was e.g.\
successfully used in \cite{e+-2}). The truth of what the correct
nonextensive thermodynamic description is will often lie inbetween
the two extreme cases (\ref{dense}) and (\ref{dilute}).

Let us now come to our main physical example, namely fully
developed turbulence. Let $u$ in
eq.~(\ref{there}) represent a local velocity difference in a
fully developed turbulent flow as measured on a certain 
scale $r$. We define
\begin{equation}
\beta= \epsilon \tau
\end{equation}
where $\epsilon$ is the (fluctuating) energy dissipation rate
(averaged over $r^3$) and $\tau$ is a
typical time scale during which energy is transferred. Both
$\epsilon$ {\em and} $\tau$ can fluctuate, and we assume that
$\epsilon \tau $ is $\chi^2$-distributed. For power-law friction forces 
the SDE (\ref{there}) generates the stationary pdf (\ref{pu}).
In Fig.~1 this theoretical distribution is
compared with experimental measurements in two turbulence
experiments, performed on two very different scales. All
distributions have been rescaled to variance 1. Apparently, there
is very good coincidence between experimental and theoretical
curves, thus indicating that our simple model assumptions are a
good approximation of the true turbulent statistics.

Generally, in turbulent systems
the entropic index $q$ 
is observed to decrease
with increasing $r$ (see \cite{BLS} for precision measurements
of $q(r)$).
At the smallest scale we expect from the definition
of the energy dissipation rate
\begin{equation}
\epsilon =5\nu\left\{ \left( \frac{\partial v}{\partial x_1}\right)^2
+\left( \frac{\partial v}{\partial x_2}\right)^2+
\left( \frac{\partial v}{\partial
x_3}\right)^2\right\}
\end{equation}
($\nu$: kinematic viscosity, $v$: velocity) that the independent
Gaussian random variables $X_i$ in eq.~(\ref{Gauss}) are given by
\begin{equation}
X_i=\sqrt{5\nu\tau} \frac{\partial v}{\partial x_i},
\end{equation}
and that there are indeed 3 of them, due to the 3 spatial
dimensions. This means $n=3$ or, using eq.~(\ref{qn}), $q\approx
\frac{3}{2}$ if $\alpha \approx 1$. This is
indeed confirmed by the fit of the small-scale data of the
Bodenschatz group in Fig.~1, yielding $q=1.49$ and an $\alpha$
as given by eq.~(\ref{qn}).

\subsection*{Acknowledgement}
I am very grateful to Harry Swinney and Eberhard Bodenschatz for
providing me with the experimental data displayed in Fig.~1.

\newpage

{\bf Fig.~1} Histogram of longitudinal velocity differences as
measured by Swinney et al. \cite{BLS, swinney} in a turbulent
Couette Taylor flow with Reynolds number $R_\lambda=262$ at scale
$r=116\eta$ (solid line), where $\eta$ is the Kolmogorov length.
The experimental data are very well fitted by the analytic formula 
(\ref{pu}) with $q=1.10$ and $\alpha=0.90$
(dashed line). The square data points are a histogram of the
acceleration (= velocity difference on a very small time scale) of
a Lagrangian test particle as measured by Bodenschatz et al. for
$R_\lambda=200$ \cite{boden}. These data are well fitted by
(\ref{pu}) with $q=1.49$ and $\alpha=0.92$ (dotted line).

\vspace{1cm}

\epsfig{file=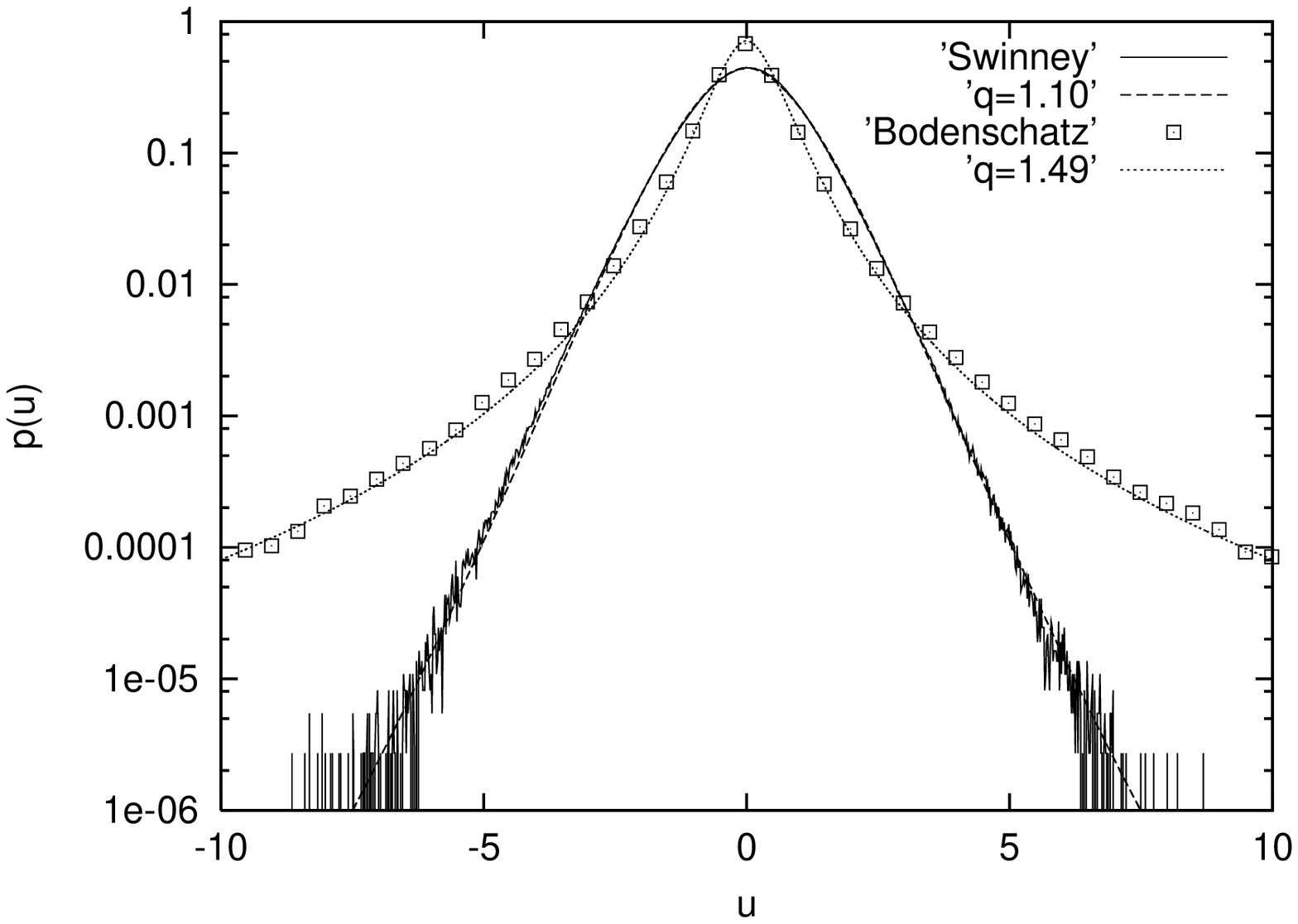, width=8cm, height=10cm, angle=-90.}
\end{document}